\newif\iffull
\fulltrue
\newif\ifcmts
\cmtstrue

\iffull
\date{Rev. 10/III/12 JM}
\else
\date{}
\fi

\title{Higher-order Erd\H{o}s--Szekeres theorems
\iffull\else\\{\large(Extended abstract)}\fi}

\ifcmts
\newcommand{\cmt}[1]{\ifhmode\newline\fi{\sf *** \ \ #1 \\}}
\else
\newcommand{\cmt}{}
\fi

\author{
{\sc Marek Eli\'a\v{s}}\thanks{Supported
by the  ERC Advanced Grant No.~267165.}\\
{\footnotesize Department of Applied Mathematics}\\[-1.5mm]
   {\footnotesize  Charles University, Malostransk\'{e} n\'{a}m. 25}\\[-1.5mm]
{\footnotesize  118~00~~Praha~1,
   Czech Republic}\\[-1.5mm]
{\footnotesize e-mail:
\href{mailto:eliam6am@ss1000.ms.mff.cuni.cz}{\nolinkurl{eliam6am@ss1000.ms.mff.cuni.cz}}}
\and
{\sc Ji\v{r}\'{\i} Matou\v{s}ek}\footnotemark[1]
\\
   {\footnotesize Department of Applied Mathematics and}\\[-1.5mm]
   {\footnotesize Institute of Theoretical Computer Science (ITI)}\\[-1.5mm]
   {\footnotesize  Charles University, Malostransk\'{e} n\'{a}m. 25}\\[-1.5mm]
{\footnotesize  118~00~~Praha~1,
   Czech Republic, and}\\
{\footnotesize    Institute of  Theoretical Computer Science}\\[-1.5mm]
{\footnotesize    ETH Zurich,
      8092 Zurich, Switzerland}\\[-1.5mm]   {\footnotesize e-mail:
      \href{mailto:matousek@kam.mff.cuni.cz}{\nolinkurl{matousek@kam.mff.cuni.cz}}}
}

\documentclass[11pt]{article}

\usepackage{a4wide}
\usepackage{latexsym}
\usepackage{amsmath}
\usepackage{amssymb}
\usepackage{graphicx}
\usepackage{url}
\usepackage{hyperref}
\hypersetup{colorlinks=true}
\hypersetup{linkcolor=black,anchorcolor=black,citecolor=black,urlcolor=black}
\hypersetup{pdftitle=Higher-order Erdos--Szekeres theorems}
\hypersetup{pdfauthor={Marek Elias} {Jiri Matousek}}
\hypersetup{pdfkeywords={Erdos-Szekeres theorem} {Ramsey theory} {hypergraph}
{ordertype} {divided difference} {k-th-order monotone subset}}

\newtheorem{theorem}{Theorem}[section]

\newtheorem{prop}[theorem]{Proposition}

\newtheorem{lemma}[theorem]{Lemma}
\newtheorem{corol}[theorem]{Corollary}

\DeclareMathOperator{\ES}{ES}
\newcommand\OT{\mathop {\rm OT}\nolimits}
\newcommand\OSH{\mathop {\rm OSH}\nolimits}

\newcommand\divdiff{\mbox{$\Delta\!\!\!\!\!\;\raisebox{0.5ex}{\mbox{\boldmath$\scriptscriptstyle\mid$}}\,\,$}}
\newcommand\twr
{\mathop {\rm twr}\nolimits}
\newcommand\rams{R}
\newcommand\trrams{\rams^{\rm trans}}

\newcommand{\heading}[1]{\vspace{1ex}\par\noindent{\bf\boldmath #1}}

\makeatletter
\newcommand{\ProofEndBox}{{\ifhmode\unskip\nobreak\hfil\penalty50 \else
          \leavevmode\fi\quad\vadjust{}\nobreak\hfill$\Box$
            \finalhyphendemerits=0 \par}}
\makeatother
\newcommand{\proofend}{\ProofEndBox\smallskip}

\newcommand{\R}{{\mathbb{R}}}

\newcommand\eps{\varepsilon}

\newcommand{\sgn}{\mathop {\rm sgn}\nolimits}

\def\:{\colon}

\long\def\onefigure#1#2{
\begin{figure*}[tbp]
\begin{center}
#1
\end{center}
\caption{#2}
\end{figure*}
}

\def\immediateFigure#1{%
\smallskip\begin{center}#1\end{center}\smallskip }

\newcommand{\labfig}[2]  
{\onefigure{\mbox{\includegraphics{high-es-#1}}}{\label{f:#1} #2} }

\newcommand{\labfigw}[3]  
{\onefigure{\mbox{\includegraphics[width=#2]{high-es-#1}}}{\label{f:#1} #3}}

\newcommand{\immfig}[1]  
{\immediateFigure{\mbox{\includegraphics{high-es-#1}}}}

\newcommand{\immfigw}[2] 
{\immediateFigure{\mbox{\includegraphics[width=#2]{high-es-#1}}}}

\begin{document}

\maketitle

\begin{abstract}
Let $P=(p_1,p_2,\ldots,p_N)$ be a sequence of points in the plane,
where $p_i=(x_i,y_i)$ and $x_1<x_2<\cdots<x_N$.
A famous 1935 Erd\H{o}s--Szekeres theorem 
asserts that every such $P$ contains a monotone subsequence $S$ of 
$\lceil\sqrt N\,\rceil$ points. Another,
  equally famous theorem from the same paper implies
that every such $P$ contains a convex or concave
subsequence of $\Omega(\log N)$ points.

Monotonicity is a property determined by pairs of points,
and convexity concerns triples of points.
We propose a generalization making both of these
theorems members of an infinite family of Ramsey-type results.
First we define a $(k+1)$-tuple $K\subseteq P$ to
be \emph{positive} if it lies on the graph of
a function whose $k$th derivative is everywhere nonnegative,
and similarly for a \emph{negative} $(k+1)$-tuple.
Then we say that $S\subseteq P$ is 
 \emph{$k$th-order monotone}
if its $(k+1)$-tuples are  all positive or all negative.
 
We investigate quantitative
bound for the corresponding Ramsey-type result
(i.e., how large $k$th-order monotone subsequence can be guaranteed
in every $N$-point~$P$). 
We obtain an $\Omega(\log ^{(k-1)}N)$
lower bound ($(k-1)$-times iterated logarithm).
This is based on a quantitative Ramsey-type theorem
for \emph{transitive colorings} of the complete
$(k+1)$-uniform hypergraph (these were recently considered
by Pach, Fox, Sudakov, and Suk).


For $k=3$, we construct a geometric example providing an
$O(\log\log N)$ upper bound, tight up to a multiplicative constant.
As a consequence, we obtain similar upper bounds for
a Ramsey-type theorem for \emph{order-type homogeneous} subsets in $\R^3$,
as well as for a Ramsey-type theorem for hyperplanes
in $\R^4$ recently used by Dujmovi\'c and Langerman.

\end{abstract}

\section{Introduction}

In this paper we mainly consider sets $P=\{p_1,p_2,\ldots,p_N\}$
of points in the plane, where $p_i=(x_i,y_i)$. 
We always assume that no two of the $x$-coordinates
coincide, and unless stated otherwise, we also
assume that the $p_i$ are numbered so that 
$x_1<x_2<\cdots<x_N$ (the same also applies to
subsets of $P$, which we will enumerate in the order
of increasing $x$-coordinates).

\heading{Two theorems of Erd\H{o}s and Szekeres. }
Among simple results in combinatorics, only few can compete
with the following one in beauty and usefulness:

\begin{theorem}[Erd\H{o}s--Szekeres on monotone
subsequences \cite{es-cpg-35}]\label{t:es1}
For every positive integer $n$,
among every $N=(n-1)^2+1$ points $p_1,\ldots,p_N\in\R^2$ as above,
one can always choose a \emph{monotone subset}
of at least $n$ points, i.e.,
indices $i_1<i_2<\cdots<i_n$ such that
either $y_{i_1}\le y_{i_2}\le \cdots\le y_{i_n}$
or $y_{i_1}\ge y_{i_2}\ge \cdots\ge y_{i_n}$.
\end{theorem}

See, for example, Steele \cite{steele-surv} for a collection 
of six nice proofs and some applications. 
\iffull
For many purposes, it is more natural to view the above
theorem as a purely combinatorial result about permutations,
but here we prefer the geometric formulation (which is
also similar to the one in the original Erd\H{o}s--Szekeres
paper).
\fi

Another result of the same paper of Erd\H{o}s and Szekeres
is the following well-known gem in discrete 
geometry:\iffull\footnote{Somewhat
unfortunately, the name Erd\H{o}s--Szekeres theorem
refers to Theorem~\ref{t:es1} in some sources
and to Theorem~\ref{t:es2} or similar statements in other sources.}\fi

\begin{theorem}[Erd\H{o}s--Szekeres on convex/concave
configurations \cite{es-cpg-35}]\label{t:es2}
For every positive integer $n$,
among every $N={2n-4\choose n-2}+1\approx 4^n/\sqrt n$ 
points $p_1,\ldots,p_N\in\R^2$ as above,
one can always choose a \emph{convex configuration}
or a \emph{concave configuration}
of  $n$ points, i.e.,
indices $i_1<i_2<\cdots<i_n$ such that
the slopes of the segments
$p_{i_j}p_{i_{j+1}}$, $j=1,2,\ldots,n-1$,
are either monotone nondecreasing or monotone nonincreasing.
\end{theorem}

See, e.g., \cite{MorrisSoltan,Mat-dg} for proofs and surveys
of developments around this result.

\heading{$k$-general position.}
To simplify our forthcoming discussion, at some places it
will be convenient to assume that the considered
point sets are in a ``sufficiently general'' position. Namely,
we define a set $P$ to be  in \emph{$k$-general position} if
no $k+1$ points of $P$ lie on the graph of a polynomial of
degree at most $k-1$. In particular,
$1$-general position requires that no two $y$-coordinates
coincide, and  $2$-general position means
 the usual general position, i.e., no three points collinear.

\heading{$k$th-order monotone subsets. }
Here we propose a view of Theorems~\ref{t:es1} and~\ref{t:es2}
as the first two members in an infinite sequence of
Ramsey-type results about planar point sets.\footnote{There is also
a (trivial) 0th member, namely, the statement that in every $P$,
at least half of the points either have  all $y$-coordinates
nonnegative or have
or all $y$-coordinates nonpositive.}

%


In Theorem~\ref{t:es1}, monotonicity of a subset
is a property of \emph{pairs} of points of the subset,
and actually, it suffices to look at pairs of consecutive points.
Similarly, convexity or concavity of a configuration 
in Theorem~\ref{t:es2} is a property
of triples, and again it is enough to look at consecutive triples.

In the former case, we are considering
the slope of the segment determined by a pair of points, 
which can be thought of as the first derivative. 
In the latter case, a triple
is convex iff its points lie on the graph of a smooth convex function,
i.e., one with nonnegative second derivative everywhere.

With this point of view, it is natural to define
a $(k+1)$-tuple $K\subseteq P$ to be \emph{positive}
if it lies on the graph of a function whose $k$-th derivative
(exists and) is everywhere nonnegative, and similarly
for a \emph{negative} $(k+1)$-tuple (in Section~\ref{s:defs},
we will provide several other, equivalent characterizations
of these properties). 
Then we say that an arbitrary subset $S\subseteq P$ is 
\emph{$k$th-order monotone} if 
 its $(k+1)$-tuples are all positive or all negative.

First-order monotonicity is obviously equivalent to monotonicity
as in Theorem~\ref{t:es1}, and second-order monotonicity
is equivalent to convexity/concavity as in Theorem~\ref{t:es2}.
We will also see (Lemma~\ref{l:trans}) that, to certify
$k$th-order monotonicity, it is enough to consider 
all $(k+1)$-tuples of \emph{consecutive} points.

Let us  remark that 
every $(k+1)$-tuple $K$ is positive or  negative,
and moreover, if $K$ is in $k$-general position, it 
cannot be both positive and negative
(Corollary~\ref{c:posneg}).
We will write $\sgn(K)=+1$
if $K$ is positive and $\sgn(K)=-1$
if $K$ is negative.

\heading{Ramsey's theorem, quantitative bounds,
and transitive colorings. } Using the just mentioned
facts, one can immediately derive a Ramsey-type theorem
for $k$th-order monotone subsets from
Ramsey's theorem.

\begin{prop}\label{p:kmonot}
For every $k$ and $n$ there exists $N$ such
that every $N$-point planar set in $k$-general position contains 
an $n$-point $k$th-order monotone subset.
\end{prop}

\heading{Proof. }
We recall Ramsey's theorem (for two colors; see, e.g.,
Graham, Rothschild, and Spencer \cite{grs-rt-90}):
for every $\ell$ and $n$ there exists $N$
such that for every coloring of the set ${X\choose \ell}$
of all $\ell$-element subsets of an $N$-element set $X$ 
there exists an $n$-element \emph{homogeneous} set $Y\subseteq X$,
i.e., a subset in which all $\ell$-tuples have the same color.
The smallest $N$ for which the claim holds is usually
denoted by $\rams_\ell(n)$. 

In our case, we set $X=P$ and color each $(k+1)$-tuple $K\subseteq P$
with the color $\sgn(K)\in\{\pm 1\}$.
Then homogeneous
subsets are exactly $k$th-order monotone subsets.
\proofend

\medskip

Let us denote by $\ES_k(n)$ the smallest value of
$N$ for which the claim in this proposition holds.
We have $\ES_1(n)\le (n-1)^2+1$ and
$\ES_2(n)\le {2n-4\choose n-2}+1$ according to
Theorems~\ref{t:es1} and~\ref{t:es2}, respectively;
moreover, these inequalities actually hold with
equality~\cite{es-cpg-35}. Our main goal
is to estimate the order of magnitude of $\ES_k(n)$ for
$k\ge 3$.

The above proof gives $\ES_k(n)\le \rams_{k+1}(n)$.
However, for $k=1$, and most likely for all $k$, the order
of magnitude of $\rams_{k+1}(n)$ is much larger than that of $\ES_k(n)$.
Indeed, considering $k$ fixed and $n$ large,
the best known lower and upper bounds of $\rams_{k+1}(n)$ are
of the form\footnote{We employ the usual asymptotic notation
for comparing functions: $f(n)=O(g(n))$ means that
$|f(n)|\le C|g(n)|$ for some $C$ and all $n$, where $C$ may depend
on parameters declared as constants (in our case on $k$);
$f(n)=\Omega(g(n))$ is equivalent to $g(n)=O(f(n))$;
and $f(n)=\Theta(g(n))$ means that both $f(n)=O(g(n))$
and $f(n)=\Omega(g(n))$.}
 $\rams_2(n)=2^{\Theta(n)}$ and, for $k\ge 2$,
\iffull
$$
\twr_k(\Omega(n^2))\le \rams_{k+1}(n) \le \twr_{k+1}(O(n)),
$$
\fi
where the tower function $\twr_k(x)$ is defined by $\twr_1(x) = x$
 and $\twr_{i+1} (x) = 2^{\twr_i (x)}$. It is widely believed
that the upper bound is essentially the truth. This belief
is supported by known bounds for more than two colors,
where the lower bound for $(k+1)$-tuples is also a tower
of height $k+1$; see Conlon, Fox, and
Sudakov \cite{conlon-al} for a recent improvement
and more detailed overview of the known bounds.

The coloring of the $(k+1)$-tuples in the above proof
of Proposition~\ref{p:kmonot} is not arbitrary. In particular,
it has a property we call \emph{transitivity} (see Lemma~\ref{l:trans}).
Transitive
colorings were introduced earlier in the recent preprint
Fox et al.~\cite[Section~6]{FoxPachSudSuk}, under the same name.

To define a transitive coloring in general, we need to consider
a hypergraph whose vertex set is linearly ordered;
w.l.o.g. we can identify it with the set $[N]:=\{1,2,\ldots,N\}$.
A coloring $c\:{[N]\choose \ell}\to [m]$ is \emph{transitive}
if, for every $i_1,\ldots,i_{\ell+1}\in [N]$,
$i_1<\cdots<i_{\ell+1}$, whenever the $\ell$-tuples $\{i_1,\ldots,i_\ell\}$
and $\{i_2,\ldots,i_{\ell+1}\}$ have the same color, then
\emph{all} $\ell$-element subsets of $\{i_1,\ldots,i_{\ell+1}\}$ have
the same color. Let $\trrams_\ell(n)$ denote the Ramsey
number for transitive colorings, i.e., the smallest $N$
such that any transitive coloring of the complete $\ell$-uniform
hypergraph on $[N]$ contains an $n$-element homogeneous subset.
We have the following bound.\footnote{By inspecting the proof of
the next theorem, it is easy to verify that the transitivity condition
is not used in full strength---it suffices to assume only that
the subsets obtained by omitting one of $i_2$, $i_3$ have the same color.}

\begin{theorem}\label{t:ub} 
For $k=1,2$, we have $\trrams_{k+1}(n)=\ES_k(n)$,
and for every fixed $k\ge 3$, 
$$
\ES_k(n)\le \trrams_{k+1}(n)\le \twr_{k}(O(n)).
$$
\end{theorem}

We note that Fox et al.~\cite{FoxPachSudSuk} proved
the slightly weaker upper bound $\trrams_{k+1}(n)\le \twr_{k}(O(n\log n))$.

\iffull
The proof of Theorem~\ref{t:ub} is given in Section~\ref{s:ub}.
\fi
The inequality $\ES_k(n)\le \trrams_{k+1}(n)$ is clear
(since every $N$-point set in $k$-general position
provides a transitive coloring of $[N]\choose k+1$).
The upper bounds for $\trrams_2(n)$
and $\trrams_3(n)$ follow by translating the proofs
of Theorem~\ref{t:es1} and~\ref{t:es2} to the setting
of transitive colorings almost word by word, and 
they are contained in \cite{FoxPachSudSuk}. 
The upper bound on $\trrams_{k+1}(n)$
is then obtained by induction on $k$, with $k=3$ as the base case,
\iffull
following one of the usual proofs of Ramsey's theorem.
\fi

\heading{A set with no large third-order monotone subsets.}
For $k\le 2$, the numbers $\ES_k(n)$ (and thus $\trrams_{k+1}(n)$)
are known exactly. Our perhaps most interesting result is
an asymptotically matching lower bound for $\ES_3(n)$.

\begin{theorem}\label{t:lb} For all $n\ge 2$
we have $\trrams_{4}(2n+1)\ge \ES_3(2n+1)\ge 2^{2^{n-1}}+1$.
Consequently, $\ES_3(n)=2^{2^{\Theta(n)}}$.
\end{theorem}

The proof is given in Section~\ref{s:lb}.
A Ramsey function with known doubly exponential growth seems to be 
rare in geometric Ramsey-type problems (a notable example
is a result of Valtr~\cite{Valtr04opencaps}).

\heading{Order types. } Here we
change the setting from the plane to $\R^d$
and we consider an ordered sequence  $P=(p_1,p_2,\ldots,p_N)$ 
in $\R^d$. This time we do \emph{not} assume the first
coordinates to be increasing.
For simplicity, we
assume $P$ to be in general position, which now means that no
$d+1$ points of $P$ lie on a common hyperplane.

We recall that \emph{order type} of $P$
specifies the orientation of every $(d+1)$-tuple of points of $P$,
and it this way, it describes purely combinatorially many of the geometric
properties of $P$. More formally, the order type of $P$
is the mapping $\chi\:{[N]\choose d+1}\to\{-1,+1\}$,
where for a $(d+1)$-tuple $I=\{i_1,\ldots,i_{d+1}\}$,
$i_1<i_2<\cdots<i_{d+1}$, $\chi(I):=\sgn\det 
M(p_{i_1},p_{i_2},\ldots,p_{i_{d+1}})$, where
$M(q_1,\ldots,q_{d+1})$ is the $(d+1)\times (d+1)$ matrix whose $j$th column
is $(1,q_j)$, i.e., $1$ followed by the vector of the $d$ coordinates of
$q_j$.
 See, e.g., 
Goodman and Pollack \cite{gp-asotd-93} or \cite{Mat-dg}
for more background about order types.

From  Ramsey's theorem for $(d+1)$-tuples, we can immediately
derive a Ramsey-type result for order types:
for every $d$ and $n$ there exists $N$ such that  every
$N$-point sequence contains an $n$-point subsequence
in which all the $(d+1)$-tuples have the same orientation
(we call such a subsequence  \emph{order-type homogeneous}).
Let us write $\OT_d(n)$ for the smallest such~$N$.

In Section~\ref{s:order} we first observe that, by simple
and probably well known considerations,
$\OT_1(n)=(n-1)^2+1$ and $\OT_2(n)=2^{\Theta(n)}$.
For $d\ge 3$, the best upper bound for $\OT_d(n)$ we are aware
of is the one from the Ramsey argument above, 
i.e., $\OT_d(n)\le\rams_{d+1}(n)\le \twr_{d+1}(O(n))$. 
In particular, for $\OT_3(n)$ this upper bound is triply
exponential; in Section~\ref{s:order} we prove
a doubly exponential lower bound.

\begin{prop}\label{p:ES<OT}
For all $d$ and $n$,  $\OT_d(n)\ge \ES_d(n)$.
In particular, $\OT_3(n)=2^{2^{\Omega(n)}}$.
\end{prop}

\heading{A Ramsey-type result for hyperplanes. } Let us consider a 
finite set $H$
of hyperplanes in $\R^d$ in general position (every $d$ intersecting
at a single point). Let us say that $H$ is \emph{one-sided} if
 $V(H)$, the vertex set of the arrangement of $H$,
lies completely on one side of the coordinate
hyperplane $x_d=0$.

Let $\OSH_d(n)$ be the smallest $N$ such that every set $H$ of
$N$ hyperplanes in $\R^d$ in general position contains a one-sided
subset of $n$ hyperplanes. Ramsey's theorem for $d$-tuples immediately
gives $\OSH_d(n)\le\rams_d(n)$ (a $d$-tuple gets color $+1$ if its intersection
has a positive last coordinate, and color $-1$ otherwise).

Matou\v{s}ek and Welzl \cite{mw-gscpt-92} observed that, actually,
$\OSH_2(n)=\ES_1(n)=(n-1)^2+1$, and applied this in a range-searching
algorithm. 
Recently Dujmovi\'c and Langerman \cite{DujLang} used the existence
of $\OSH_d(n)$ (essentially Lemma~9 in the arXiv version of their paper)
to prove several interesting results, such as a ham-sandwich
and centerpoint theorems for hyperplanes. 
\iffull 

In Section~\ref{s:order}
 we show \fi
that lower bounds for $k$th-order monotone subsets
in the plane can be translated into lower bounds 
for~$\OSH_d$.

\begin{prop}\label{p:onesided} We have $\OSH_d(n)\ge \ES_{d-1}(n)$,
and in particular, $\OSH_3(n)=2^{\Omega(n)}$ and\footnote{An exponential
lower bound for $\OSH_3$
was known to the authors of \cite{mw-gscpt-92},
and perhaps to others as well, but as far as we know, it hasn't
appeared in print.} 
$\OSH_4(n)=2^{2^{\Omega(n)}}$.
\end{prop}

\iffull
The lower bounds for $\OSH_d(n)$ can also be translated into lower
bounds in the theorems of Dujmovi\'c and Langerman. For example,
in their ham-sandwich theorem, we have $d$ collections $H_1,\ldots,H_d$
of hyperplanes in $\R^d$, each of size $N$, and we want a hyperplane
$g$ such that in each $H_i$, we can find disjoint subsets $A_i,B_i$ of $n$
hyperplanes each such $V(A_i)$ lies on one side of $g$ and
$V(B_i)$ on the other side. 

To derive a lower bound for
the smallest necessary $N$, we fix $d$ affinely independent points
$p_1,\ldots,p_d$ in the $x_d=0$ hyperplane, and a set $H$ of $N$ hyperplanes
in general position with no one-sided subset of size~$n$.
We let $H_i$ be an affinely transformed copy of $H$
 such that all of $V(H_i)$ lies very close to~$p_i$. 
Then every potential ham-sandwich hyperplane $g$ for these $H_i$ has to be 
almost parallel to the $x_d=0$ hyperplane, and thus there cannot
be $A_i,B_i$ of size~$n$ for all $i$.
\fi 

\heading{The work of Fox et al. } While preparing
a draft of the present paper, we learned about a recent preprint
of Fox, Pach, Sudakov, and Suk \cite{FoxPachSudSuk}.
They investigated various combinatorial and geometric
problems  inspired by Theorems~\ref{t:es1} and~\ref{t:es2},
and as was mentioned above, among others, they 
introduced transitive colorings,\footnote{With still another geometric
source of such colorings besides the Erd\H{o}s--Szekeres
theorems, namely, noncrossing convex bodies in the plane}
but mainly they studied a related but different
Ramsey-type quantity:
let $N_\ell(q,n)$ be the smallest integer $N$ such that,
for every coloring of ${[N]\choose \ell}$ with $q$ colors,
there exists an $n$-element $I=\{i_1,\ldots,i_n\}\subseteq [N]$,
$i_1<\cdots<i_n$,
inducing a \emph{monochromatic monotone path},
i.e., such that all the $\ell$-tuples 
of the form $\{i_j,i_{j+1},\ldots,i_{j+\ell-1}\}$, $j=1,2,\ldots,n-\ell+1$,
have the same color. 

They note that $\trrams_\ell(n)\le N_\ell(2,n)$, and they
obtained the following bounds for $N_\ell(2,n)$:
$N_2(2,n)=\ES_1(n)$, $N_3(2,n)=\ES_2(n)$, and for every fixed $k\ge 3$,
\iffull
$$
\twr_{k}(\Omega(n))\le N_{k+1} (2, n)\le \twr_{k}(O(n\log n)).
$$
\fi
As we mentioned after Theorem~\ref{t:ub}, this also yields
an upper bound for $\trrams_{k+1}(n)$ 
only slightly weaker than the one in that theorem.

\heading{Open problems. } 
\begin{enumerate}
\item
We have obtained reasonably tight  bounds
for $\ES_3(n)$, but the gaps 
are much more significant for $\ES_k(n)$ with $k\ge 4$.
According to the cases $k=1,2,3$, one may guess that
$\ES_k(n)$ is of order $\twr_k(\Theta(n))$,
and thus that stronger lower bounds are needed,
but a possibility of a better upper bound shouldn't also be overlooked.
This question looks both interesting
and challenging.
\item
A perhaps more manageable task might
be a better lower bound for $\trrams_k(n)$, $k\ge 4$.
A natural approach would be to imitate the Stepping-Up Lemma
used for lower bounds for the Ramsey numbers $\rams_k(n)$
(see, e.g., \cite{conlon-al}). But
so far we have not succeeded in this, since 
even if we start with a transitive coloring of $k$-tuples,
we could not guarantee transitivity for
the coloring of $(k+1)$-tuples.
\item
As for order-type homogeneous sequences, for $\OT_3(n)$
we have the lower bound of $2^{2^{\Omega(n)}}$, but upper bound
only $\twr_4(O(n))$ directly from Ramsey's theorem. It seems that the 
colorings given by the order type are not transitive in
any reasonable sense, and we have no good guess of which
of the upper and lower bounds should be closer to the truth.
Similar comments apply to the problem with one-sided subsets
of planes in $\R^3$ (concerning $\OSH_3(n)$), and the higher-dimensional
cases are even more widely open.
\iffull
\item
Another interesting question is whether $n\log n$ can be
replaced by $n$ in the upper bound for the quantity $N_\ell(2,n)$
considered by Fox et al.~\cite{FoxPachSudSuk}.
\fi
\item
In our definition of $k$th-order positivity, every $(k+1)$-tuple
of points should lie on the graph of a function with a nonnegative
$k$th derivative, and different functions can be used for different
$(k+1)$-tuples. In an earlier version of this paper,
we conjectured  that, assuming $k$-general position,
a single function should suffice for all $(k+1)$-tuples;
in other words, that every $k$th-order monotone
finite set finite set in $k$-general position
lies on a graph of a $k$-times differentiable function $f\:\R\to\R$ whose
$k$th derivative is everywhere nonnegative or everywhere
nonpositive. 

However, Rote \cite{Rote} disproved this 
for $k=3$ (while the cases $k=1,2$ do hold, as is not hard
to check). With his kind permission, we reproduce his
example at the end of Section~\ref{s:defs}.

Naturally, this opens up interesting new questions: How can one characterize
point sets lying on the graph of a function whose $k$th derivative
is positive everywhere? Is there a Ramsey-type theorem for such sets,
and if yes, how large is the corresponding Ramsey function?
\end{enumerate}

\section{On the definition of \boldmath $k$th-order monotonicity}
\label{s:defs}

Here we provide several equivalent characterizations
of $k$th-order monotonicity of planar point sets
and some of their properties. First we recall several known
results.

\heading{Divided differences and Newton's interpolation. }
Let $p_1,p_2,\ldots,p_{k+1}$ be points in the plane, $p_i=(x_i,y_i)$,
where the $x_i$ are all distinct
(but not necessarily increasing). We recall that the
\emph{$k$th divided difference} $\divdiff_k(p_1,p_2,\ldots,p_{k+1})$ 
is defined recursively as follows:
\begin{eqnarray*}
\divdiff_0(p_1)&:=&y_1\\
\divdiff_k(p_1,p_2,\ldots,p_{k+1})&:=&
\frac{\divdiff_{k-1}(p_2,p_3,\ldots,p_{k+1})-
\divdiff_{k-1}(p_1,p_2,\ldots,p_{k})}{x_{k+1}-x_1}.
\end{eqnarray*}
For example, $\divdiff_1(p_1,p_2)$ equals the slope of the
line $p_1p_2$. In general, the $k$th divided difference
is related to the $k$th derivative as follows
(see, e.g., \cite[Eq.~1.33]{phillips}; note
that the case $k=1$ is the Mean Value Theorem):

\begin{lemma}[Cauchy]\label{l:cauchy}
Let the points $p_1,\ldots,p_{k+1}$,
$a:=x_1<x_2<\cdots<b:=x_{k+1}$, lie on the graph of a function
$f$ such that the $k$th derivative $f^{(k)}$ exists
everywhere on the interval $(a,b)$.
Then there exists $\xi\in (a,b)$ such that
$$
\divdiff_k(p_1,\ldots,p_{k+1})=\frac{f^{(k)}(\xi)}{k!}.
$$
\end{lemma}

We will also need the following result
(see, e.g., \cite[Eq.~1.11--1.19]{phillips}).

\begin{lemma}[Newton's interpolation]\label{l:newton}
Let $p_1,\ldots,p_{k+1}\in\R^2$ be points with distinct
$x$-coordinates (here we need not assume that the $x$-coordinates
are increasing). Then the unique
polynomial $f$ of degree at most $k$ whose graph contains
 $p_1,\ldots,p_{k+1}$ is given by
$$
f(x)=\sum_{i=1}^{k+1} \biggl(\divdiff(p_1, \ldots, p_{i})
 \prod_{j=1}^{i-1} (x - x_j)
\biggr) 
$$
In particular, the coefficient of $x^k$ is
$\divdiff(p_1, \ldots, p_{k+1})$, and it equals 
 $f^{(k)}(x)/k!$ (which is a constant function).
\end{lemma}

\iffull
We recall that  a $(k+1)$-tuple $K=\{p_1,\ldots,p_{k+1}\}$
was defined to be  positive if it
is contained
in the graph of a function having a nonnegative $k$th derivative
everywhere. We obtain the following equivalent characterization:
\fi

\begin{corol}\label{c:posneg} 
A $(k+1)$-tuple $K=\{p_1,\ldots,p_{k+1}\}$
is positive iff $\divdiff_k(p_1,\ldots,p_{k+1})\ge 0$
(and similarly for a negative $(k+1)$-tuple).
If $K$ is in $k$-general position, we have
$\sgn K=\sgn \divdiff_k(p_1,\ldots,p_{k+1})$.
\end{corol}

\heading{Proof. } If $K$ is contained in the graph of $f$
with $f^{(k)}\ge 0$ everywhere, then $\divdiff_k(p_1,\ldots,p_{k+1})\ge 0$
by Lemma~\ref{l:cauchy}. 

Conversely, if $\divdiff_k(p_1,\ldots,p_{k+1})\ge 0$,
then by Lemma~\ref{l:newton}, the unique polynomial
of degree at most $k$  whose graph contains
$K$ is the required function with nonnegative
$k$th derivative.

If, moreover, $K$ is in $k$-general position, then
$\divdiff_k(p_1,\ldots,p_{k+1})\ne 0$, and
so $K$ cannot be both $k$th-order positive
and $k$th-order negative by Lemma~\ref{l:cauchy}.
\proofend


We will also need the following criterion for the sign
of a $(k+1)$-tuple.

\begin{lemma}\label{l:signs}
Let $K=\{p_1,p_2,\ldots,p_{k+1}\}$ be a $(k+1)$-tuple of points
in $k$-general position, $x_1<\cdots<x_{k+1}$,
let $i\in[k+1]$,
and let $f_i$ be the (unique) polynomial of degree at most $k-1$
whose graph passes through
the points of $K\setminus\{p_i\}$. Then $\sgn K=(-1)^{k-i}$ 
if $p_i$ lies  below the graph of $f_i$,
and $\sgn K=(-1)^{k+1-i}$ if $p_i$ lies above the graph.
\end{lemma}

\long\def\proofA{Let $f$ be the polynomial of degree
at most $k$ passing through all of $K$.
We use Newton's interpolation (Lemma~\ref{l:newton}),
but with the points reordered so that $p_i$ comes last,
and we get that
$$
f(x)=f_i(x)+\divdiff_k(p_1,\ldots,p_{i-1},p_{i+1},\ldots,p_{k+1},p_i)
\prod_{j\in[k+1]\setminus\{i\}} (x-x_j).
$$
Using this  with $x=x_i$, we get 
\begin{eqnarray*}
\sgn (y_i-f_i(x_i))&=&\sgn(f(x_i)-f_i(x_i))\\
&=&
\sgn \divdiff_k(p_1,\ldots,p_{i-1},p_{i+1},\ldots,p_{k+1},p_i)
\cdot \sgn \prod_{j\in[k+1]\setminus\{i\}} (x_i-x_j).
\end{eqnarray*}

Divided differences are invariant under
permutations of the points (as can be seen, e.g., from
Lemma~\ref{l:newton}, since the interpolating polynomial
does not depend on the order of the points), and so
$\sgn \divdiff_k(p_1,\ldots,p_{i-1},p_{i+1},\ldots,p_{k+1},p_i)=
\sgn K$.  Finally, the product
 $\prod_{j\in[k+1]\setminus\{i\}} (x_i-x_j)$ has
$k+1-i$ negative factors, thus its sign is $(-1)^{k+1-i}$,
and the lemma follows.
\proofend}
\iffull\proofA \fi

It remains to prove transitivity.

\begin{lemma}\label{l:trans} Let $P=\{p_1,\ldots,p_N\}$
be a point set in $k$-general position. Then the $2$-coloring
of $(k+1)$-tuples $K\in {P\choose k+1}$ by their sign
is transitive. 
\end{lemma}

\heading{Proof. } We consider a $(k+2)$-tuple
$L=\{p_1,\ldots,p_{k+2}\}$ with $\sgn\{p_1,\ldots,p_{k+1}\}=
\sgn \{p_2,\ldots,p_{k+2}\}=+1$, and we fix $i\in\{2,\ldots,{k+1}\}$.
Let $f_{i,k+2}$ be the polynomial of degree at most $k-1$
passing through $L\setminus \{p_i,p_{k+2}\}$, and similarly
for $f_{1,k+2}$. Our goal is to show that $f_{i,k+2}(x_{k+2})<y_{k+2}$,
since this gives $\sgn(L\setminus\{p_i\})=+1$ by 
Lemma~\ref{l:signs}.

Since $\sgn(L\setminus\{p_1\})=+1$, we have
$f_{1,k+2}(x_{k+2})<y_{k+2}$ (Lemma~\ref{l:signs} again), 
and so it suffices to prove $f_{i,k+2}(x_{k+2})<f_{1,k+2}(x_{k+2})$. 

Let us consider the polynomial $g:=f_{1,k+2}-f_{i,k+2}$;
as explained above, our goal is proving $\sgn g(x_{k+2})=+1$.
To this end, we first determine $\sgn g(x_1)$:
We have $f_{i,k+2}(x_1)=y_1$ and
$\sgn(y_1-f_{1,k+2}(x_1))=(-1)^{k}$
(using $\sgn(L\setminus\{p_1\})=+1$ and Lemma~\ref{l:signs}).
Hence $\sgn g(x_1)=(-1)^{k-1}$.

Next, we observe that $g$ 
is a polynomial of degree at most $k-1$, and it vanishes
at $x_2,\ldots,x_{i-1},x_{i+1},\ldots,x_{k+1}$. These are $k-1$
distinct values; thus, they include all roots of $g$,
and each of them is a simple root. Consequently,
$g$ changes sign $(k-1)$-times between $x_1$ and~$x_{k+2}$.
Hence, finally, $\sgn g(x_{k+2})=(-1)^{k-1}\sgn g(x_1)=+1$
as claimed.
\proofend

\begin{sloppypar}
\heading{Rote's example. } Fig.~\ref{f:roteex} shows a 6-point
set $P=\{p_1,\ldots,p_6\}$ in 3-general position (no four points
on a parabola). It is easy to check 3rd-order positivity
using Lemma~\ref{l:signs}:
By transitivity, it suffices to look at $4$-tuples of consecutive
points. For $p_1,\ldots,p_4$ we use the parabola through
$p_1,p_2,p_3$ (which actually degenerates to the $x$-axis);
for $p_2,\ldots,p_5$ we use the dashed parabola through $p_2,p_3,p_4$
(which is very close
to the $x$-axis in the relevant region); and for
$p_3,\ldots,p_6$, the parabola through $p_4,p_5,p_6$ (drawn full).
\end{sloppypar}

\labfig{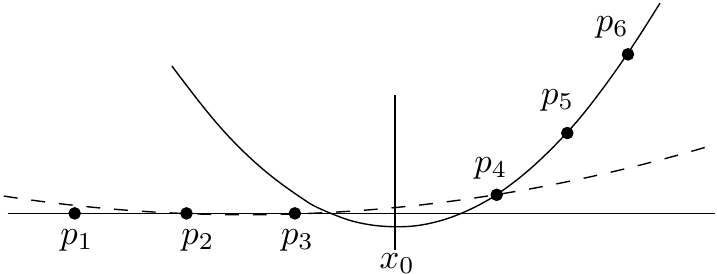}{Rote's example: a 6-point 3rd-order positive set
in 3-general position that does not lie on the graph
of any function with nonnegative 3rd derivative.}

It remains to check that $P$ does not lie on
the graph of a function $f$ with $f^{(3)}\ge 0$ everywhere.
Assuming for contradiction that there is such an $f$,
we consider the point $q:=(x_0,f(x_0))$,
where $x_0$ is such that the full parabola is below the $x$-axis
at $x_0$. For the $4$-tuple $\{p_1,p_2,p_3,q\}$ to be positive,
$q$ has to lie above the $x$-axis, but the $4$-tuple
$\{q,p_4,p_5,p_6\}$ is positive only if $q$ lies below the
parabola through  $p_4,p_5,p_6$---a contradiction.

\long\def\proofB{
As we mentioned
in the remark following that theorem, it suffices to 
establish the case $k\ge 3$. 

Thus, we want to prove that $\trrams_{k+1}(n)\le \twr_{k}(C_k n)$
for all $n$ and for every $k\ge 3$, with suitable constants $C_k$
depending on $k$. As the base of the induction we use
$\trrams_3(n)\le 4^n$, which, as was remarked earlier,
follows by imitating the proof of Theorem~\ref{t:es2}.

Thus, let $k\ge 3$ be fixed, let $n$ be given, and 
let us set $M:=\trrams_k(n)$. We will prove that
\begin{equation}\label{e:recur}
\trrams_{k+1}(n)\le N:=2^{M^{k}}.
\end{equation}
Theorem~\ref{t:ub} then follows from this recurrence and
from the fact that $2^{\twr_{k-1}(n)^k}\le \twr_{k}(k n)$
for $k\ge 3$, which is easy to check. 

To prove (\ref{e:recur}), we follow an inductive proofs
of Ramsey's theorem going back to Erd\H{o}s and Rado
\cite{erdRado}.
Let $\chi\:{[N]\choose k+1}\to\{1,2\}$ be an arbitrary transitive
2-coloring. We set $A_{k-1}:=\{1,2,\ldots,k-1\}$ and $X_{k-1}:=
[N]\setminus A_{k-1}$.
For $i=k,k+1,\ldots,M$ we will inductively construct 
sets $A_i,X_i\subseteq [N]$ such that 
\begin{enumerate}
\item[(i)] $A_i<X_i$ (i.e., all elements
of $A_i$ precede all elements of $X_i$);
\item[(ii)] $|A_i|=i$ and $|X_i|\ge |X_{i-1}|/2^{M^{k-1}}$; and
\item[(iii)] the color of a $(k+1)$-tuple whose first $k$ elements
all belong to $A_i$ does not depend on its last element;
in other words, for $K\in {A_i\choose k}$ and $x,y\in A_i\cup X_i$
with $K<\{x,y\}$, we have $\chi(K\cup\{x\})=\chi(K\cup\{y\})$.
\end{enumerate}

For the inductive step, suppose that $A_{i}$ and $X_i$ have already
been constructed. We let $x_i$ be the smallest element of $X_i$,
we set $A_{i+1}:=A_i\cup\{x_i\}$, and we write $X'_i:=X_i\setminus\{x_i\}$. 

Let us  call two elements $x,y\in X'_i$
\emph{equivalent} if we have, for every $K\in {A_{i-1}\choose k-1}$,
$\chi(K\cup\{x_i,x\})=\chi(K\cup\{x_i,y\})$.
There are ${i\choose k-1}$ possible choices of $K$,
and hence there are at most $2^{i\choose k-1}< 2^{M^{k-1}}$
equivalence classes. 
We choose $X_{i+1}\subseteq X'_i$
as the largest equivalence class. Then (i), (iii) obviously hold
for $A_{i+1}$ and $X_{i+1}$, and we have
$|X_{i+1}|\ge (|X_i|-1)/(2^{M^{k-1}}-1)\ge |X_i|/2^{M^{k-1}}$
(since $i\le M$ and thus we have $|X_{i}|\ge N/(2^{M^k-1})^{i-1}=
2^{M^{k}-(i-1)M^{k-1}}\ge 2^{M^{k-1}}$). 
This finishes the inductive construction
of $A_i$ and~$X_i$.

In this way, we construct the sets $A:=A_M$ and $X_M$
(note that $|X_M|\ge 1$ by (ii)). Let $x$ be the first element
of $X_M$, and let us define a 2-coloring
$\chi^*\:{A\choose k}\to\{1,2\}$ of the $k$-tuples of $A$
by $\chi^*(K):=\chi(K\cup\{x\})$.

We claim that, crucially, $\chi^*$ is transitive (which is not
entirely obvious). So we consider elements $a_1<a_2<\cdots<a_{k+1}$
of $A$, and we suppose that $\chi^*(\{a_1,\ldots,a_k\})=
\chi^*(\{a_2,\ldots,a_{k+1}\})=:c$. We want to show that
$\chi^*(\{a_1,\ldots,a_{k+1}\}\setminus\{a_i\})=c$ for every
$i=2,3,\ldots,k$. 
We have $c=\chi^*(\{a_1,\ldots,a_k\})=\chi(\{a_1,\ldots,a_k,x\})=
\chi(\{a_1,\ldots,a_{k+1}\})$ (by definition and by the independence
of $\chi$ of the last element), and $c=\chi^*(\{a_2,\ldots,a_{k+1}\})=
\chi(\{a_2,\ldots,a_{k+1},x\})$. Next we use the transitivity
of $\chi$ on the $(k+2)$-tuple $(a_1,\ldots,a_{k+1},x)$,
obtaining $\chi(\{a_1,\ldots,a_{k+1},x\}\setminus\{a_i\})=c=
\chi^*(\{a_1,\ldots,a_{k+1}\}\setminus\{a_i\})$ as needed.

Now we can apply the inductive hypothesis to $A$, which yields
an $n$-element subset of $A$ homogeneous w.r.t.~$\chi^*$,
and this subset is homogeneous w.r.t.~$\chi$ as well, finishing
the proof of Theorem~\ref{t:ub}.
\proofend
}

\iffull
\section{Upper bounds on the Ramsey numbers for transitive
colorings}\label{s:ub}

In this section we prove Theorem~\ref{t:ub}. \proofB

\fi

\section{\boldmath A lower bound for $\ES_3$}\label{s:lb}

Here we prove Theorem~\ref{t:lb},
a lower bound for $\ES_3(2n+1)$. 
We proceed by induction on~$n$;
the goal is to construct a set $P_n$ of $N:=2^{2^{n-1}}$ points
with no $(2n+1)$-point third-order monotone subset.
The induction starts for $n=2$ with an arbitrary 
$P_2$ of size $2^{2^1}=4$.

In the inductive step, given $P_n$, we will construct $P_{n+1}$
so that $|P_{n+1}|=|P_n|^2$; then the bound on the size
of $P_n$ clearly holds.

We may assume that $P=P_n$ is in $3$-general position (this can always
be achieved by a small perturbation). By an affine transformation
we also make sure that $P\subset [1,2]\times [0,1]$;
or actually, $P\subset [1,1.9]\times [0,1]$ so that there is
some room for perturbation.
Moreover, there is a small $\delta>0$ such that if $P'$
is obtained from $P$ by moving each point arbitrarily by
at most $\delta$, then $P'$ is still in $3$-general position,
the order of the points of $P'$ along the $x$-axis is the same
as that for $P$,
and the sign of every $4$-tuple in $P'$ is the same as the
sign of the corresponding $4$-tuple in~$P$.

\heading{The construction.} The construction of $P_{n+1}$ from
$P=P_n$ as above proceeds in the following steps.
\begin{enumerate}
\item We choose a sufficiently large number $A=A(P)$
(the requirements on it will be specified later), and
we set $\eps:=1/A^2$.
\item For every point $p\in P$, let $Q_p$ be the image of $P$ under the
affine map that sends the square $[1,2]\times[0,1]$ to the axis-parallel
rectangle of width $\eps$, height $\eps^2$, and with the lower left
corner at $p$; see Fig.~\ref{f:constr}.
\item Let $\psi_p(x)=Ax^2+C_p$ 
be a quadratic function, where  $A$ is as above and
$C_p$ is chosen so that $\psi_p(x(p))=0$
(where $x(p)$ is the $x$-coordinate of $p$).
 Let $\breve{Q}_p$ be the set
obtained by ``adding $\psi_p$ to $Q_p$'', i.e., by
shifting each point $(x,y)\in Q_p$ vertically upwards by $\psi_p(x)$.
We set $P_{n+1}:=\bigcup_{p\in P} \breve{Q}_p$. We call the
$\breve{Q}_p$ the \emph{clusters} of~$P_{n+1}$.
\end{enumerate}
\labfig{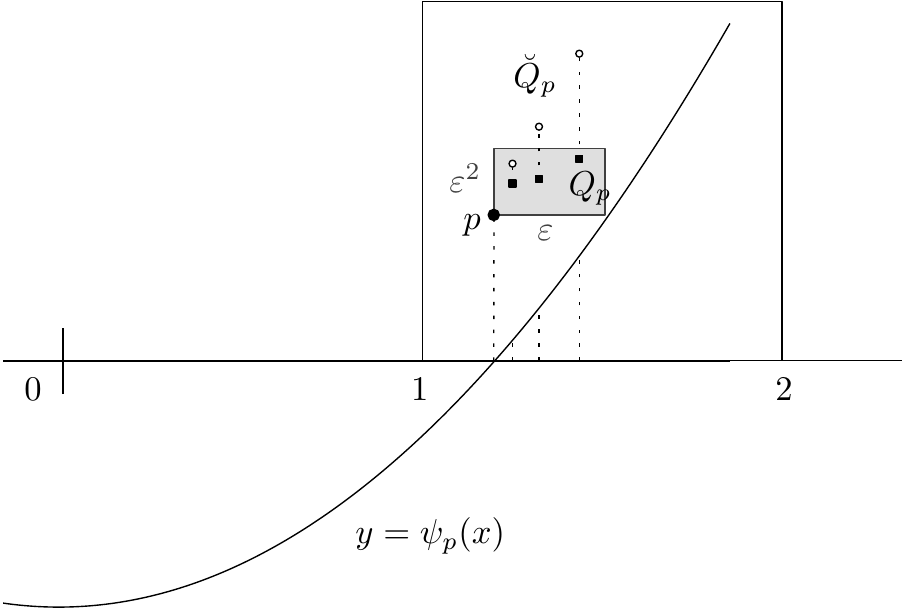}{A schematic illustration of the construction of $P_{n+1}$.}

First we check that each cluster $\breve{Q}_p$ lies close 
to~$p$.

\begin{lemma}\label{l:Zp-close}
 Each $\breve{Q}_p$ is contained in an $O(\sqrt\eps\,)$-neighborhood of~$p$.
\end{lemma}

\iffull

\heading{Proof.} Writing $p=(x_0,y_0)$, 
the set $Q_p$ obviously lies in the $2\eps$-neighborhood of $p$,
and the maximum
amount by which a point of $Q_p$ was translated upwards 
is at most 
$$
\psi_p(x_0+\eps)=A\left((x_0+\eps)^2-x_0^2\right)=A(2x_0\eps+\eps^2)=
O(\sqrt\eps\,).
$$
\proofend
\fi

Here is a key property of the construction.
 
\begin{lemma}[Slope lemma]\label{l:slope}
Let $\lambda$ be a parabola passing through three points of $P_{n+1}$
that belong to three different clusters, or a line passing through two points
of different clusters.
Let $\mu$ be a parabola passing through three points of a single 
cluster $\breve{Q}_p$,
or a line passing through two such points. Then  the maximum
slope (first derivative) of $\lambda$ on the interval $[1,2]$ is smaller than
the minimum slope of $\mu$ on $[1,2]$, 
provided that $A$ 
was chosen sufficiently large.
\end{lemma}

\heading{Proof. } Clearly, the maximum slope of any such $\lambda$ 
can be bounded from above by some finite number
depending only on $P$ but not on~$A$. Thus, it suffices to show
that, with $A$  large, for every $\mu$ as in the lemma, 
the minimum slope is bounded from below by~$A$.

First let us assume that $\mu$ is a parabola passing through three
points of $\breve{Q}_p$, where $p=(x_0,y_0)$, let $\tilde \mu$
be the parabola passing through the corresponding three points
of $P$, and let the equation of $\tilde\mu$ be $y=ax^2+bx+c$.

By the construction of $\breve{Q}_p$, the affine map transforming $P$
to $Q_p$ sends a point with coordinates $(x,y)$ to the point
$(\eps(x-1)+x_0,\eps^2 y+y_0)$. Calculation shows that
the image of $\tilde\mu$ under this affine map has
the equation  $y=a x^2+(2 a \eps + b \eps - 2 a x_0)x+
c'$, where the value of the absolute term $c'$ need not be calculated
since it doesn't matter. Hence the 
minimum slope of this curve 
on $[1,2]$ is bounded from below by $-(8|a|+4|a|\eps+2|b|\eps+8|a|)$.
Finally, $\mu$ is obtained by adding $\psi_p(x)=Ax^2+C_p$ to this curve,
and the minimum slope of $\psi_p$ on $[1,2]$ is at least~$2A$. 

Next, let $\mu$ be a line passing through two points $q,r\in \breve{Q}_p$.
Let us choose another point $s\in \breve{Q}_p$ and consider the parabola
$\mu'$ through $q,r,s$. By the Mean Value Theorem, the slope
of $\mu$ equals the slope of $\mu'$ at some point between
$q$ and $r$, and the latter is at least~$A$ by the above.
The lemma is proved.
\proofend

\medskip

Let $K=\{p_1,p_2,p_3,p_4\}\subseteq P_{n+1}$ be a 4-tuple,
$p_i=(x_i,y_i)$,
$x_1<\cdots<x_4$. We assign a \emph{type} to $K$, which
is an ordered partition of $4$ given by the distribution
of $K$ among the  clusters; for example,
$K$ has type $1+1+2$ if the first point $p_1$ lies
in some $\breve{Q}_p$, $p_2$ lies in $\breve{Q}_{p'}$ for $p'\ne p$,
and $p_3,p_4\in \breve{Q}_{p''}$, $p''\ne p,p'$.

The next lemma shows that the sign $K$ is determined by its
type. We provide a complete classification, although we will not
use all of the types in the subsequent proof.

\begin{lemma}\label{l:type-signs}
Let $K=\{p_1,p_2,p_3,p_4\}\subseteq P_{n+1}$ be a $4$-tuple. 
If $K$ is
of type $1+1+1+1$ or $4$, then the sign of $K$ is the same
as that of the corresponding $4$-tuple in $P$.
Otherwise, the
sign of $K$ is determined by its type as follows:
\begin{itemize}
\item
for types $3+1$ and $1+3$ it is $-1$;
\item
for types $1+1+2$ and $2+1+1$ it is $+1$;
\item
for type $1+2+1$ it is $-1$; and
\item
for type $2+2$ it is $+1$.
\end{itemize}
\end{lemma} 

\heading{Proof. } Since the transformation that converts $P$
into $\breve{Q}_p$ preserves the types of $4$-tuples, the statement
for type 4 is clear. The statement for type $1+1+1+1$
follows since, by Lemma~\ref{l:Zp-close}, $K$ is obtained by
a sufficiently small perturbation of the corresponding $4$-tuple in~$P$
(this gives one of the lower bounds on $A$, since
we need the bound in Lemma~\ref{l:Zp-close} to be smaller
than the $\delta$ considered at the beginning of
our description of the construction).

The statements for the remaining types are obtained by simple 
application of the slope lemma
(Lemma~\ref{l:slope}) together with Lemma~\ref{l:signs}.
Namely, for type $3+1$, we get that the parabola through $p_1,p_2,p_3$ 
lies above $p_4$ (by comparing its slope to the slope of the line
$p_3p_4$); see Fig.~\ref{f:slopes}. 
For type $1+3$ we similarly get that $p_1$ lies above the parabola
through $p_2,p_3,p_4$, and so the sign is $-1$ in both of these cases.
\labfig{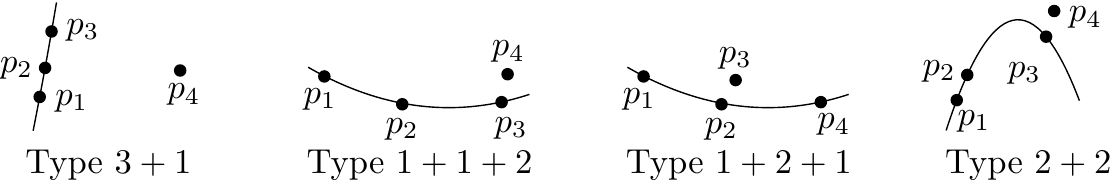}{Determining the signs of $4$-tuples by type.}

For type $1+1+2$, the segment $p_3p_4$ is steeper than the parabola
through $p_1p_2p_3$, and so the sign is $+1$. Similarly for type
$2+1+1$ we get that $p_1$ lies below the parabola through $p_2,p_3,p_4$,
which again gives sign $+1$. For type $1+2+1$, $p_3$ lies above
the parabola through $p_1,p_2,p_4$, giving sign $-1$. 
Finally, for type $2+2$, the segment $p_1p_2$ is steeper than
$p_2p_3$, thus the parabola through $p_1,p_2,p_3$ is concave,
and hence its slope at $p_3$ and after it is no larger than
the slope of the segment $p_2p_3$. Thus, $p_4$ lies above this
parabola and the sign is $+1$ as claimed.
\proofend

\heading{Finishing the proof of Theorem~\ref{t:lb}. } 
It remains to show that $P_{n+1}$ contains no $(2n+3)$-point
third-order monotone subset.

For contradiction, suppose that $M\subseteq P_{n+1}$
is such a $(2n+3)$-point subset. Let $2n+3=
n_1+n_2+\cdots+n_s$ be the type of $M$ (i.e., $M$ has
$n_i\ge 1$ points in the $i$th leftmost cluster it intersects). 
By the inductive assumption we have $s\le 2n$ and $n_i\le 2n$
for all $i$. 

Let $n_a=\max_i n_i$ and $n_b=\max_{i\ne a} n_i$ be the two
largest among the $n_i$. 
For convenience, let us assume $a<b$; 
the case $a>b$ is handled symmetrically.
We distinguish three cases.

First, if $n_a\ge 3$ and $n_b\ge 2$, then we can select
4-tuples  of types $3+1$ and $2+2$ from the corresponding two clusters,
which have different signs, and so $M$ is not homogeneous.

Second, if $n_a\ge 3$ and $n_b=1$, then we have at least three $n_i$
equal to 1 (since $n_a\le 2n$), and at least two of them lie
on the same side of the cluster corresponding to $n_a$,
say to the right of it. Then we can select 4-tuples of
types $3+1$ and $2+1+1$, again of opposite signs.

Third, if $n_a=2$, then there are at least two other clusters
of size 2. From these three 2-element clusters, we can select
4-tuples of types 2+2 and 1+2+1, again of opposite signs.

This exhausts all possibilities ($n_a=1$ cannot happen,
because $s\le 2n$), and Theorem~\ref{t:lb} is proved.
\proofend

\section{Order types and one-sided sets of hyperplanes}
\label{s:order}

First we substantiate the two claims made above Proposition~\ref{p:ES<OT},
concerning $\OT_1$ and $\OT_2$.
For $d=1$, an order-type homogeneous sequence in $\R^1$ is just a
monotone sequence of real numbers, so $\OT_1(n)=(n-1)^2+1$
by Theorem~\ref{t:es1}. 

In a similar spirit, it is easy to check that
a planar order-type homogeneous sequence 
corresponds to the vertices of a convex $n$-gon, 
enumerated in a clockwise or counterclockwise
order. Thus, $\OT_2(n)\ge \ES_2(\lceil n/2\rceil)=
2^{\Omega(n)}$. On the other hand,
given any $N$-point sequence, we can first select
a subsequence of $\lceil \sqrt N\,\rceil$ points with
increasing or decreasing $x$-coordinates, and then we
select a convex or concave
configuration from it. Thus, by Theorem~\ref{t:es2},
we have $\OT_2(n)=2^{O(n)}$.

\heading{Proof of Proposition~\ref{p:ES<OT}. }
For a point $p=(x,y)\in\R^2$, we define
the point $\tilde p:=(x,x^2,\ldots,x^{d-1},y)\in\R^d$.

To prove that $\ES_d(n)\le \OT_d(n)$, we consider
a set $P=\{p_1,\ldots,p_N\}\subset \R^2$ in $d$-general position, 
$p_i=(x_i,y_i)$,
where $N=\ES_d(n)-1$ and
$x_1<\cdots<x_N$,
with no $d$th-order monotone subset of $n$ points.
It suffices to prove 
 that the sequence $\tilde P:=(\tilde p_1,\tilde p_2,\ldots,\tilde p_N)$
has no $n$-point order-type homogeneous subsequence.  This follows
from the next lemma.

\begin{lemma}\label{l:vanderm} 
For every $(d+1)$-tuple $(p_1,\ldots,p_{d+1})$ of points in $\R^2$,
$x_1<\cdots<x_{d+1}$, we have
$\sgn(\{p_1,\ldots,p_{d+1}\})=
\sgn \det M(\tilde p_1,\tilde p_2,\ldots,\tilde p_{d+1})$,
where $M(q_1,\ldots,q_{d+1})$ is the matrix from the definition of 
order type above Proposition~\ref{p:ES<OT}.
\end{lemma}

\heading{Proof. }  By Lemma~\ref{l:newton}
and Corollary~\ref{c:posneg}, the sign
of $\{p_1,\ldots,p_{d+1}\}$
 equals the sign of the coefficient $a_d$ of the
unique polynomial  $f(x)=\sum_{j=0}^d a_j x^j$ of degree at most $d$
whose graph passes through the points $p_1,\ldots,p_{d+1}$.

The vector $a=(a_0,\ldots,a_d)$ can be expressed
as the solution of the linear system $Va=y$, where $y=(y_1,\ldots,y_{d+1})$
and $V$ is the \emph{Vandermonde matrix} with $v_{ij}=x_i^{j-1}$,
$i,j=1,2,\ldots,d+1$. By Cramer's rule, we obtain
$$a_d=\frac{\det W}{\det V},
$$
where $W$ stands for the matrix $V$ with the last column
replaced with the vector $y$. As is well known,
$\det V=\prod_{1\le i<j\le d+1}(x_j-x_i)$, and since
$x_1<\cdots<x_{d+1}$, we have $\det V>0$. 
Thus, $\sgn a_d=\sgn \det W$. Finally, we have
$$
W=\left(\begin{array}{cccccc}
1&x_1&x_1^2&\ldots&x_1^{d-1}&y_1\\
\vdots&\vdots&\vdots&\vdots&\vdots&\vdots\\
1&x_{d+1}&x_{d+1}^2&\ldots&x_{d+1}^{d-1}&y_{d+1}
\end{array}
\right)=M(\tilde p_1,\tilde p_2,\ldots,\tilde p_{d+1})^T.
$$
The lemma follows, and Proposition~\ref{p:ES<OT} is proved.
\proofend

\long\def\proofC{
\heading{Proof of Proposition~\ref{p:onesided}. }
The proof is very similar to the 
\iffull previous one. 
\fi 
This time we start with 
a set $P=\{p_1,\ldots,p_N\}\subset \R^2$ in $(d-1)$-general position,
$p_i=(x_i,y_i)$,
where $N=\ES_{d-1}(n)-1$ and
$x_1<\cdots<x_N$, with no $(d-1)$th-order monotone 
subset of $n$ points. We define a collection $H=\{h_1,\ldots,h_N\}$
of $N$ hyperplanes in $\R^d$, where $h_i$ is given by 
$$
h_i=\biggl\{(\xi_1,\ldots,\xi_d)\in\R^d:\sum_{j=1}^d x_i^{j-1} \xi_j=y_i\biggr\}.
$$
The intersection point $\xi=(\xi_1,\ldots,\xi_d)$
of, say, $h_1,\ldots,h_d$ is the solution of the linear system
$V\xi=y$, where $V$ is the $d\times d$ Vandermonde matrix this time,
$v_{ij}=x_i^{j-1}$. Cramer's rule then gives that the $d$th coordinate $\xi_d$,
whose sign we are interested in, equals $(\det W)/(\det V)$,
where $W$ is obtained from $V$ by replacing the last column with~$y$.

As we saw in the proof of Proposition~\ref{p:ES<OT}, $(\det W)/(\det V)$
also expresses the leading coefficient in the polynomial
of degree $d-1$ passing through $p_1,\ldots,p_d$, and thus its sign
equals $\sgn \divdiff_{d-1}(p_1,\ldots,p_d)$. It follows that
one-sided subsets of $H$ precisely correspond to $(d-1)$st-order monotone
subsets in $P$, and the proposition is proved.
\proofend
}

\iffull\proofC\fi

\iffull\section*{Acknowledgment}\fi
We would like to thank J\'anos Pach for kindly discussing some of the
results of Fox et al.~\cite{FoxPachSudSuk} with us. We also thank
G\"unter Rote for informing us about about his refutation of our
conjecture and for permission to present it in this paper.

\bibliographystyle{alpha}
\bibliography{cg,geom}

\end{document}